\documentclass{amsart}
\usepackage{amsaddr}

\RequirePackage{amsthm,amsmath,amsfonts,amssymb, float, graphicx, bm, bbm, subfig, multirow}

\usepackage{color}
\usepackage[justification=centering]{caption}
\usepackage[round]{natbib}
\allowdisplaybreaks

\newcommand{\X}{\bm{X}}
\newcommand{\Z}{\bm{Z}}
\newcommand{\Y}{\bm{Y}}
\newcommand{\E}{\bm{E}}
\newcommand{\V}{\textbf{var}}
\newcommand{\PP}{\mathbb{P}}

\newcommand{\bdot}{ \boldsymbol{\cdot}}

\newtheorem{theorem}{Theorem}[section]

\newtheorem{lemma}[theorem]{Lemma}

\title[Graph-based Inference for Random Effects]{Graph-theoretic Inference for Random Effects in High-Dimensional Studies}
\author{Lynna Chu and Yichuan Bai}
\email{lchu@iastate.edu; ycbai@iastate.edu}
\address{Department of Statistics, Iowa State University}

\graphicspath{{/Users/lchu/Library/CloudStorage/Dropbox/gbRandomEffects/Draft/Figures/}}

\begin{document}

\begin{abstract}
We study the problem of testing for the presence of random effects in mixed models with high-dimensional fixed effects. To this end, we propose a rank-based graph-theoretic approach to test whether a collection of random effects is zero. Our approach is non-parametric and model-free in the sense that we not require correct specification of the mixed model nor estimation of unknown parameters. Instead, the test statistic evaluates whether incorporating group-level correlation meaningfully improves the ability of a potentially high-dimensional covariate vector $X$ to predict a response variable $Y$. We establish the consistency of the proposed test and derive its asymptotic null distribution. Through simulation studies and a real data application, we demonstrate the practical effectiveness of the proposed test. 

\end{abstract}
\maketitle

\section{Introduction}
Correlated data structures, such as longitudinal and multilevel data, arise naturally across a wide range of scientific disciplines, including genetics, economics, agriculture, education, and public health.  Mixed models provide a natural extension of regression models that allow practitioners to account for such correlations, which may result from group-level clustering (i.e. students within the same classroom) or repeated measures. These models offer a framework to accommodate correlations among the observations by incorporating both fixed effects ($\bm{X})$ and unobserved random effects ($u$), enabling a broad class of correlation structures to be explicitly specified. 

In this work, we focus on developing a statistical inference procedure to test whether a collection of random effects is zero; that is, whether the inclusion of a random effect is supported by the data at hand. Ignoring relevant random effects can lead to inflated Type I error or reduced power when testing fixed effects \citep{cui_what_2016}. We target the setting where the fixed effects can be high-dimensional ($p>N$) while the random effects are low-dimensional. In the high-dimensional setting, it can be inherently challenging to correctly specify a model, especially in the presence of a possible random effect. The relationship between $\Y$ and the covariates may be linear (with respect to the parameters) or governed by complex, non-linear mechanisms, further complicating model specification. Robust tools for diagnosing misspecification or validating model fit are limited in the high-dimensional setting and additional assumptions, such as sparsity in the true model, are often needed \citep{muller2013model}. 

To address these challenges, we develop a non-parametric testing procedure that does not rely on correct specification of the mixed model or estimation of unknown parameters. Our approach is model-free in the sense that it leverages similarity information among observations without imposing strong assumptions on the relationship between $\Y$, $\bm{X}$, and $u$. We construct a test statistic based on a graph-theoretic framework and rank-transformed data, which together capture the extent to which accounting for group-level correlation improves the predictive relationship between $\bm{X}$ and $\Y$. This framework offers robustness to model misspecification, while the graph-based structure enables a wide range of dependence patterns to be considered.

Inference for random effects has been well studied in low-dimensional settings. Recent contributions include \cite{jiang_linear_2021}, \cite{ekvall_confidence_2022}, and references therein. \cite{lee2012permutation} introduced permutation tests for random effects, demonstrating their validity and power in small sample sizes. However, inference for the existence of random effects in high-dimensional settings remains an open and challenging problem. Most existing works assume a linear mixed model. For example,  \cite{chen_inference_2015} considered ANOVA-type designs, while \cite{li2022inference} discuss inference for fixed effects and estimation of variance components in high-dimensional linear mixed models.  \cite{law_inference_2023} proposed an asymptotic F-test for testing random effects in high-dimensional linear mixed models under sparsity assumptions; their approach requires estimating the fixed effects regression coefficients via an exponential weighting procedure. From a model selection perspective, \cite{schelldorfer2011estimation} proposed an $l_1$ penalty procedure for model selection and estimation in high-dimensional linear mixed models, with extensions developed in \cite{wang2012penalized} and \cite{groll2014variable}. \cite{fan2012variable} proposed a variable selection strategy for both fixed and random effects in linear mixed models, allowing the number of fixed effects to grow exponentially with the sample size when the cluster sizes are balanced. More recently, \cite{yang2023model} proposed a clustering method for high-dimensional longitudinal data that simultaneously performs clustering and variable selection.

\subsection{Notation and assumptions}

In what follows, we refer to correlated group data as clustered data. Let $s=1, \ldots, I$ denote the cluster indices. Let $C_s$ denote the indices of the observations in cluster $s$,  $n_s = |C_s|$ be the number of observations in cluster $s$, and $N = \sum_{s=1}^I n_s$ be the total number of observations. 
We denote the unobserved random vector $u \in \mathbb{R}^I$ as the random effects, which are assumed to have mean zero and finite variance-covariance matrix $\Psi = \textbf{cov}(u)$. Let $\varepsilon \in \mathbb{R}^N$ denote the vector of random errors, which captures any variability not accounted for by the fixed or random effects. We assume that all elements of $\varepsilon$ are uncorrelated with the elements of $u$ and that $\varepsilon$ are independently distributed with mean zero and variance $\sigma^2_\varepsilon$. 
In our setup, we have $(X_i, Y_i), i = 1, \hdots, N$ of observed ordered pairs, such that the response variable $Y_i \in  \mathbb{R}$ takes on continuous values and $X_i = (X_{i1}, \hdots, X_{ip}) \in \mathbb{R}^p$ is a $p$-dimensional vector of fixed effects measured for observation $i$. Collectively, we denote the $N \times p$ matrix corresponding to the fixed effects, $\X$, and the $N$-dimensional vector of responses, $\Y.$ We do not impose any assumptions on the distribution of $\Y$ nor do we specify a parametric relationship between $\Y$ and $\X$. 



\section{Inference for Random Effects in High-dimensional Settings} \label{sec:method}

\subsection{New Test Statistic}

We are interested in testing $H_0: \Psi=0$ versus $H_1: \Psi \neq 0$. Under $H_0$, we assume that there are no random effects, and that any systematic variability in $\Y$ can be attributed to $\X$, while the remaining variability is due to random noise. We want to discern whether incorporating additional correlation structures, i.e. introducing a random effect, can significantly improve the ability of $\X$ to predict $\Y$.

To address this, we consider a generalized correlation coefficient proposed by \cite{friedman_graph-theoretic_1983} which extends the notion of correlation to multivariate observations. In particular, the authors proposed an asymmetric measure of association that evaluates how well a vector $X_i$ can be used to make single-valued prediction of a variable $Y_i$, $i = 1, \hdots, N$, without regard to how well $X_i$ can be predicted from $Y_i$. 

To test $H_0$ against $H_1$, we adapt the test statistic from \cite{friedman_graph-theoretic_1983} and propose a new graph-theoretic test statistic. Specifically, let $G_{\bm{X}}$ be a similarity graph constructed over the interpoint distances between the $\bm{X}$ observations, where each observation $X_i$ is a node in the graph, and each node pair defines an edge. The similarity graph can be constructed according to a specific criteria; examples include a $k$-minimum spanning tree ($k$-MST) or $k$-nearest neighbor graph ($k$-NN) using Euclidean distance. In practice, any similarity measure that can reasonably be defined on the sample space can be used. Then, for each observation $Y_i$, we rank the other observations $Y_j \, (j \neq i)$  in increasing order of their distance from $i$ in the $\bm{Y}$ space. Let $R_i(j)$ denote the normalized rank of observation $Y_j$ with respect to $Y_i$, defined as $R_i(j) = (r_{ij}-1)/N$, where $r_{ij}$ is the raw rank of $Y_j$ based on its distance to $Y_i$. Notice that the ranks are not symmetric: in general $R_i(j) \neq R_j(i)$. 
Define 
\begin{align*}
a_{ij} & = \begin{cases} 1 & \text{ if edge } (i,j) \in G_{\X}, \\
0 & \text{ otherwise, }
\end{cases} \\
& \text{ and } \\
c_{ij}(s) & = \begin{cases} 1 & \text{ if } (X_i,X_j) \in C_s, \\
0 & \text{ otherwise }. 
\end{cases} 
\end{align*}

\begin{figure}[H]
\centering
\includegraphics[width=0.7\linewidth]{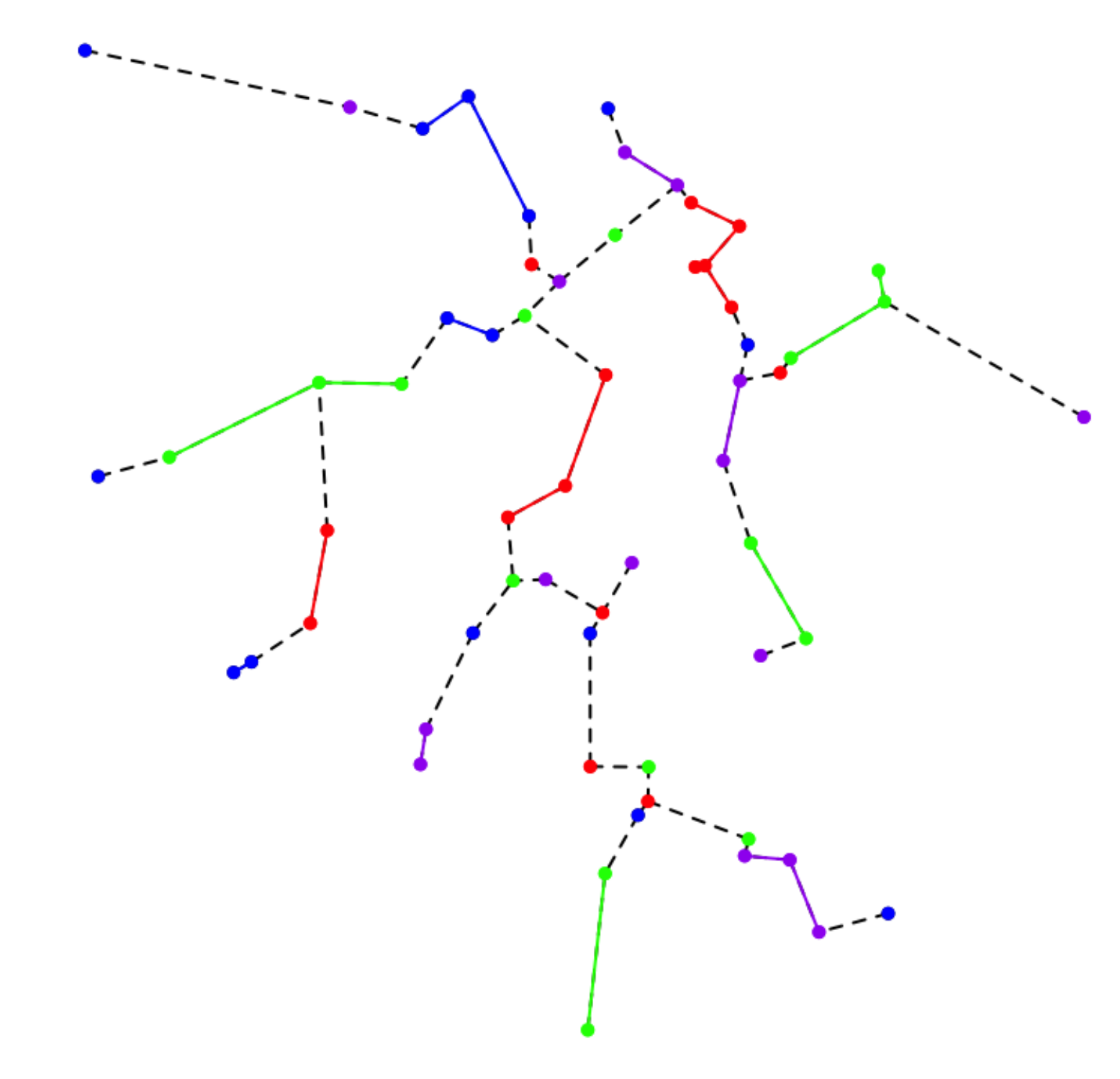}
\caption{$G_{\X}$ constructed on the $\bm{X} \in \mathbb{R}^{N \times p}$ space, such that $N = 60$, $s = 1, \hdots, 4$, and $p =2$ . Here $G_{\X}$ is an MST using Euclidean distance; each $X_i$ is a node in the graph and its color represents its cluster label. The solid lines represent edges connecting within clusters. Blue solid edges correspond to $C_1$, purple solid edges to $C_2$, red solid edges to $C_3$, and green solid edges to $C_4$. The dashed black lines represent edges connecting between clusters.}
\label{fig:Gx4} 
\end{figure}

Then the following statistic 
\begin{align*}
 T_s & = \sum_{i=1}^N \sum_{i=1}^N a_{ij} R_{i}(j) c_{ij}(s)\\
 & = \sum_{(i,j) \in G_x}  R_{i}(j) \mathbbm{1}\{(X_i, X_j) \in C_s \}
 \end{align*}
 is the \textit{within-cluster edge-rank}, since only the ranks of observations that have an edge in $G_{\bm{X}}$ and belong to cluster $s$ contribute to the test statistic. An example of the similarity graph $G_{\X}$ is illustrated in Figure \ref{fig:Gx4}. 

To test whether or not there is evidence for a random effect, we propose the following test statistic: 
\begin{equation}
Z_I =  \frac{\mathcal{V}_I- \E(\mathcal{V}_I)}{\sqrt{\V(\mathcal{V}_I)}}, 
\end{equation}
where $\mathcal{V}_I := \sum_{s=1}^I T_s$. 
Rejection of $H_0$ occurs for small values of $Z_I$. The rationale is if there is within-cluster correlation, then observations from the same cluster ($c_{ij}(s) = 1$) should tend to be close in $\bm{X}$ (i.e., have an edge $(i,j) \in G_{\X}$), and should also be close in $\bm{Y}$ space, which corresponds to relatively small values of $R_i(j)$. On the other hand, if there is insufficient evidence to reject the null hypothesis, then the cluster labels are non-informative. Since there is an association between $\bm{Y}$ and $\bm{X}$, we would still expect observations that are close in $\bm{X}$ to also be close in $\bm{Y}$, but not as strongly as under the alternative hypothesis.  This rationale holds even if the signal-to-noise ratio between $\bm{Y}$ and $\bm{X}$ is relatively small.

\begin{figure}[!ht]
\centering
\subfloat[Boxplots of $T_s$ for $I=4$ clusters.]{\label{exp:ts1}
\centering
\includegraphics[width=0.5\linewidth]{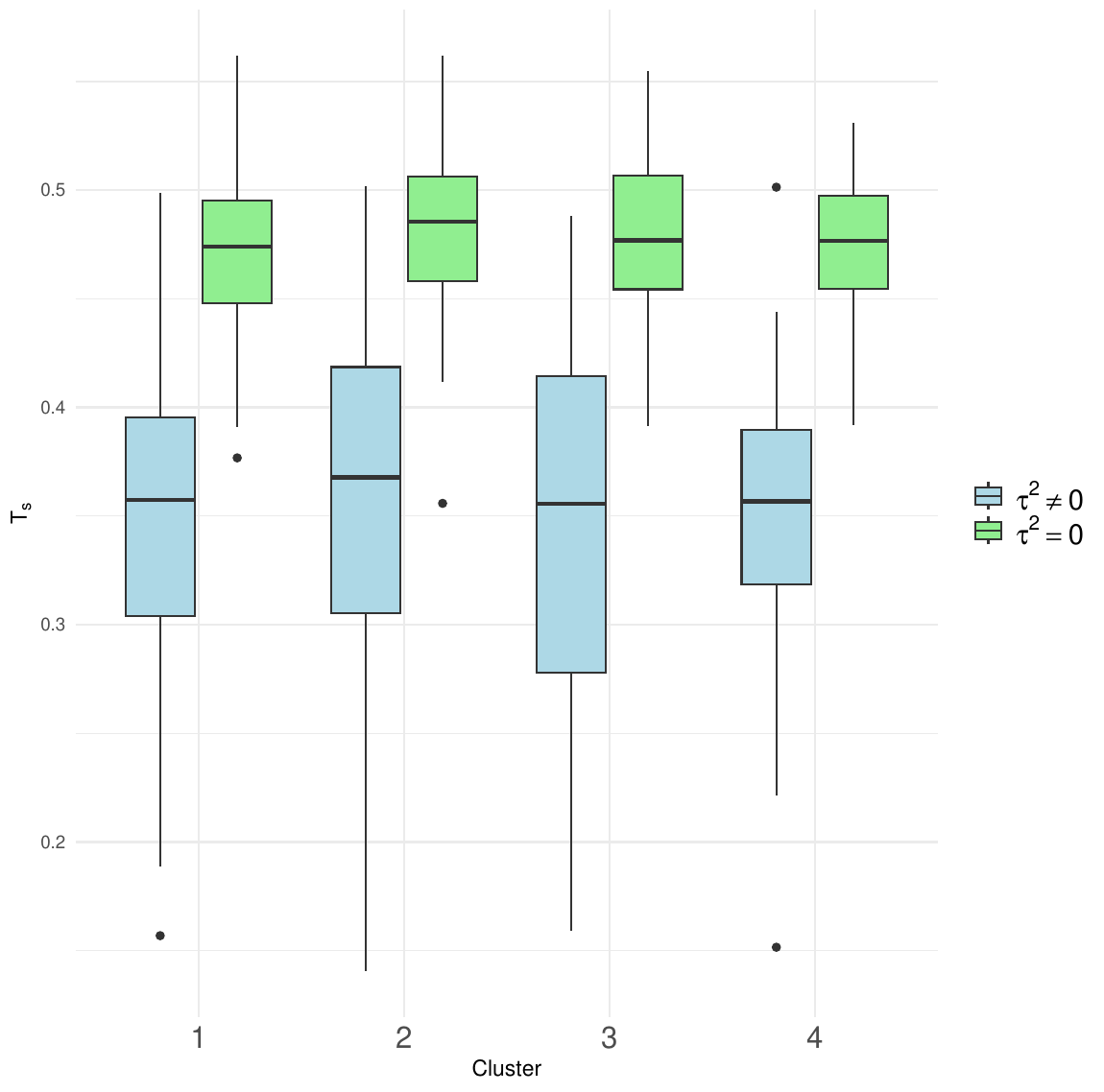}
}
\subfloat[Boxplots of $\sum_{s=1}^I T_s$ for $I=4$ clusters.]{\label{exp:ts2}
\centering
\includegraphics[width=0.5\linewidth]{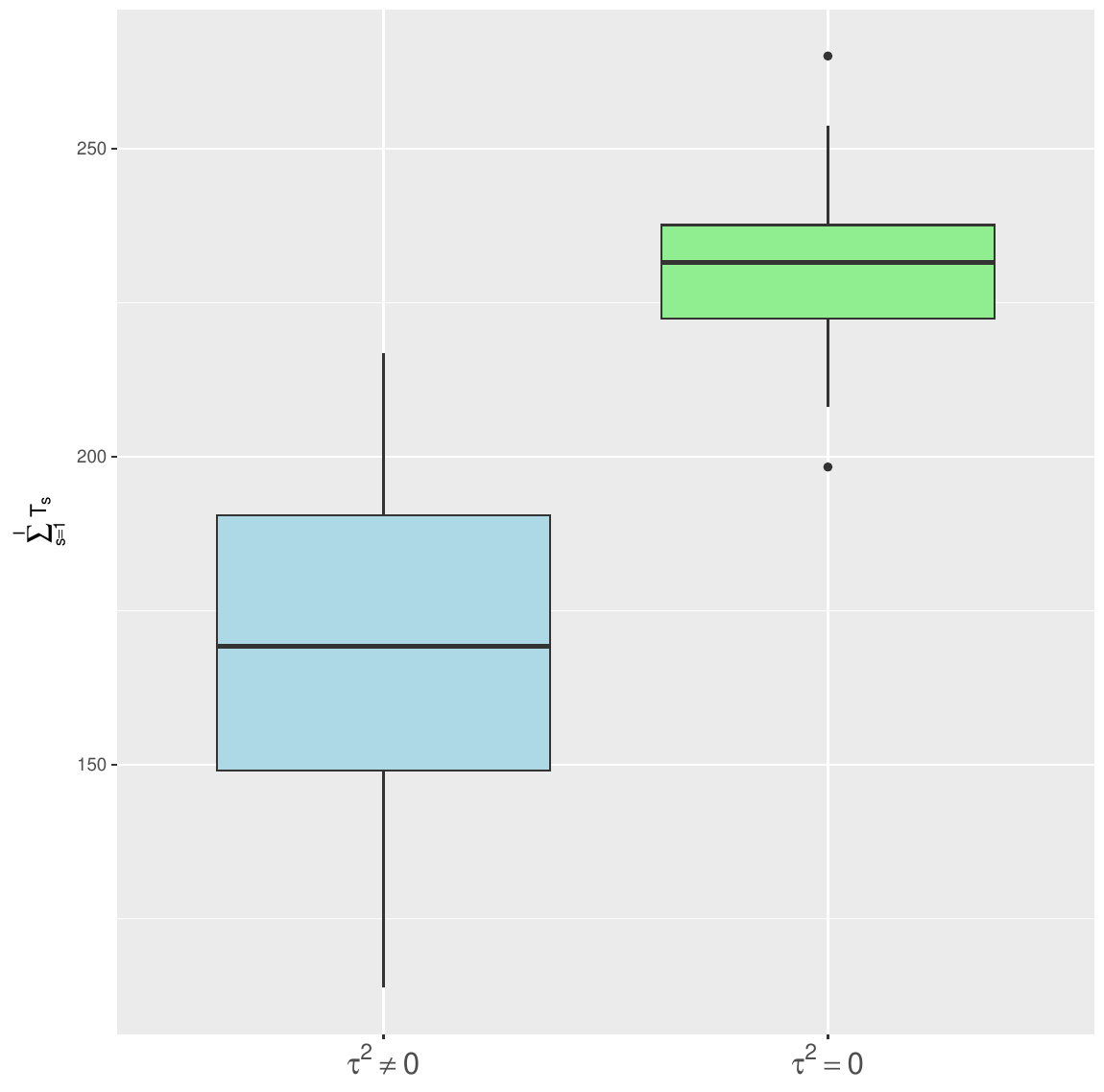}
}
\caption{Boxplots of $T_s$ and $\sum_{s=1}^I T_s$ for four clusters under the null ($\tau^2 = 0$) and alternative ($\tau^2 \neq 0$) hypothesis over 100 trials.}
\label{exp:ts}
\end{figure}

We use the following example to illustrate the difference in the behavior of $T_s$ under the null and alternative hypotheses. Consider the linear mixed model $\Y = \X\beta + \bm{Z}u + \varepsilon$, with $\X \in \mathbb{R}^{N \times p}$, $\Y, \varepsilon \in \mathbb{R}^N$, $\bm{Z} \in \mathbb{R}^{N \times I}$ and $u \in \mathbb{R}^I$.  Let $N = 200$, $p =200$, and $I= 4$. The design matrix $Z$ has a balanced one-way ANOVA design. The random effects and random errors are normally distributed, and we set $\Psi = \tau^2 \mathbf{I}$. We examine the behavior of within-cluster edge-ranks under two settings: the null setting ($\tau^2 = 0$) and the alternative setting ($\tau^2 = 4$). The similarity graph $G_{\X}$ is the MST constructed using Euclidean distance. If we have evidence to reject $H_0$, we expect $T_s$ to be smaller than it would be under the null hypothesis. Figure \ref{exp:ts} presents boxplots of $T_s$ and $\sum_{s=1}^I T_s$, $s = 1, \hdots, 4$. We can see that under the alternative ($\tau^2 = 4$), the within-cluster edge-ranks are smaller compared to the null setting. 

Under $H_0$, there is no random effect and the cluster labels are exchangeable, with equal probability given to all possible ${N \choose n_1, n_2, \hdots, n_I}$ shuffling of the cluster labels. We refer to this as the permutation null distribution. If the cluster labels are indeed exchangeable, then our observed test statistic should be no more extreme than the distribution of the statistic calculated under permutation.  Since the graph $G_{\X}$ is determined by the values of $X_i's$, not the cluster labels, it remains constant under permutation. Similarly, the ranks of $Y_i$ also remain constant under permutation. When there is no further specification, we denote $\PP$, $\E$, and $\V$ as the probability, expectation, and variance under the permutation null distribution. While our test statistic builds on \cite{friedman_graph-theoretic_1983}, the null and alternative hypotheses we consider are distinct, and the permutation procedure is conducted over cluster labels. Consequently, this necessitates new theoretical derivations to characterize the analytical expressions under the permutation null distribution. 

The analytical expressions for the expectation and variance of $T_s$ can be obtained via combinatorial analysis and are presented in the following lemma. 
\begin{lemma}\label{app:exp}
\begin{align*}
\E(T_s)  =  & \, \frac{n_s (n_s - 1)}{N(N-1)}R_{\bm{X}},\\
\V(T_s)  = & \, \frac{n_s(n_s-1)}{N(N-1)} \left\{ R_{\bm{X},1} + R_{\bm{X},2} + \frac{(n_s -2)}{(N-2)} \left( \frac{(N-n_s)}{(N-3)}  \sum_{(i,j) \in G_{\X}} R_i(j) \bar{R}_{ij}  \right. \right. \\
& \left. \left. + \frac{(n_s - 2N+3)}{(N-3)} (R_{\X,1} + R_{\X,2}) + \frac{(n_s-3)}{(N-3)} R^2_{\X} \right) - \frac{n_s(n_s-1)}{N(N-1)} R^2_{\X} \right\}, \\
\textbf{cov}(T_s ,T_t) 
 = & \, \frac{n_s n_t (n_s-1)(n_t-1)}{N(N-1)(N-2)(N-3)} \left( \frac{(4N-6)}{N(N-1)} R^2_{\X} + R_{\X,1} + R_{\X,2} - \sum_{(i,j) \in G_{\X}} R_i(j) \bar{R}_{ij} \right),
\end{align*}
where $R_{\bm{X}} = \sum_{(i,j) \in G_{\bm{X}}} R_i(j)$, $R_{\X,1} = \sum_{(i,j) \in G_{\X}} R^2_i(j)$, $R_{\X,2} = \sum_{(i,j) \in G_{\X}}R_i(j) R_j(i) $,\\ $\bar{R}_{ij} = (R_{i \bdot} + R_{\bdot i} + R_{\bdot j} + R_{j \bdot})$ with  $R_{i \bdot} = \sum_{(i,v) \in G_{\X}} R_i(v)$ 
and $R_{\bdot i} = \sum_{(i,v) \in G_{\X}} R_v(i). $ 
\end{lemma}

The proof to this lemma is provided in Appendix A.1. 

It follows that 
\begin{align*}
\E(\mathcal{V}_I) = \E\left(\sum_{s=1}^I T_s\right) & = \sum_{s=1}^I \frac{n_s (n_s - 1)}{N(N-1)}  R_{\X},  \\
\V(\mathcal{V}_I) = \V \left(\sum_{s=1}^I T_s\right) 
& = \sum_{s=1} \V(T_s) + 2\sum_{s<t} \textbf{Cov}(T_s,T_t). 
\end{align*}

\section{Theoretical Results} \label{sec:theory}

\subsection{Asymptotic Null Distribution} 
As the amount of data increases, so does the number of possible permutations, which quickly makes exact enumeration of all possible permutations computationally unfeasible. While an approximate permutation distribution can be generated through Monte Carlo sampling, this can also become computationally expensive as the sample size increases. To facilitate fast and convenient implementation of the test, we derive its asymptotic null distribution. 

Before presenting the theorem, we define two additional quantities of the similarity graph $G_{\X}$. Let $(i,j)$ represent an edge in $G_{\bm{X}}$ connecting observations $X_i$ and $X_j$. We define
\begin{align*}
&A_{(i, j)} = \{(i, j)\} \cup \begin{aligned}[t]\{&(i', j')\in G_{\bm{X}} : (i', j') \text{ and } (i, j) \text{ share a node}\},\end{aligned}\\
&B_{(i, j)} = A_{(i, j)} \cup\begin{aligned}[t]
\{&(i'', j'')\in G_{\bm{X}}: \exists (i', j')\in A_{(i, j)}, \\ & \text{ s.t. } (i', j')\text{ and } (i'', j'') \text{ share a node}\}. 
\end{aligned}
\end{align*}

Then, $A_{(i,j)}$ is the subgraph of $G_{\bm{X}}$ consisting of all edges that share a node with edge $(i,j)$. Similarly, $B_{(i,j)}$ is the subgraph in $G_{\bm{X}}$ consisting of all edges that share a node with any edge in $A_{(i,j)}$. Let $R_{B_{(i,j)}} := \sum_{(i', j')\in B_{(i, j)}}R_{i'}(j')$ and $R_{A_{(i,j)}} := \sum_{(i', j')\in A_{(i, j)}}R_{i'}(j')$. 

\begin{theorem}\label{thm:limit_null}
Under conditions
\begin{enumerate}
    \item $|G_{\X}| = \mathcal{O}(N^{\alpha}), 1\leq\alpha<1.5$,
    \item $ R_{A_{(i,j)}} = O(| A_{(i, j)}|R_{i}(j))$, $\forall (i, j) \in G_{\X}$,
    \item $\sum\limits_{(i, j)\in G_{\X}}(R_{i}(j)|A_{(i, j)}|)^2 = o(R_{\X,1} N^{0.5})$,
    \item $\sum\limits_{(i, j)\in G_{\X}}R_i(j)R_{A_{(i,j)}} R_{B_{(i,j)}} =o(R_{\X,1}^{1.5})$
\end{enumerate} 

 and under the permutation null distribution, as $n_s$, $N\rightarrow\infty$ with $n_s/N \rightarrow p_s \in (0, 1), \forall s \in [1, \dots, I]$, such that $\sum_s^I p_s =1$, we have $$Z_I \xrightarrow{\mathcal{D}} \mathcal{N}(0,1). $$
\end{theorem}

This implies, under $H_0$, that the test with rejection region $$\{ Z_I \le z_{\alpha}\}$$ is asymptotically level $\alpha$, where $z_\alpha$ is the $\alpha$th quantile of the standard normal distribution. 

Details of the proof are provided in Appendix A.2. 
The conditions are sufficient to ensure that the test statistic is well-behaved. Condition (1) places a constraint on density of $G_{\bm{X}}$; for a $k$-MST this density is controlled by the choice of $k$.  
Condition (2) ensures that observations that are close in $\bm{X}$ (share an edge) cannot be too far apart in terms of ranks in $\bm{Y}$ space. Conditions (3) and (4) ensure that the sum of ranks in the subgraph is bounded and concentrated. 

To assess the validity of the asymptotic results for finite sample sizes, we conduct simulation studies to evaluate whether the asymptotic $p$-values adequately control the empirical test size and how they compare to the permutation $p$-values. The specific simulation setup and results are described in Section \ref{sec:null}. 

\subsection{Consistency}

A test is said to be consistent if its power converges to $1$ under fixed alternatives. To establish this, we assume that, after integrating out the random effect, the marginal distributions of $Y_s$ have densities $f_s, s = 1, \hdots I$, such that $\forall s \neq t$, $f_s \neq f_t$ differ on a set of positive Lebesgue measure. Furthermore, under the alternative hypothesis, we assume that the within-cluster ranks are bounded above by a constant $\gamma_s > 0$ for each cluster $s$; that is, $R_i(j) \le \gamma_s$ for all $i,j \in C_s$. This condition ensures that the within-cluster ranks cannot be too large under the alternative, reflecting sufficient within-cluster similarity. The consistency result is stated below. 

\begin{theorem} \label{thm:consistency}
Suppose $k = O(1)$ and as $N \rightarrow \infty$, $n_s/N \rightarrow p_s \in (0,1)$, for all $s= 1, \hdots, I$. Then, under all fixed alternatives for which there exists constants $\gamma_s >0$ such that the within-cluster ranks in cluster $s$ are bounded above by $\gamma_s$, the test based on $Z_I$ is consistent. 
\end{theorem} 

Details of the proof are provided in Appendix A.3. 

\section{Simulation Studies} \label{sec:sim}

\subsection{Calibration Under the Null Hypothesis} \label{sec:null} 

To show that our test is well-calibrated under the null hypothesis of no random effect, we simulate data from a linear mixed model:
\begin{equation}\label{eq:lmm}
    \bm{Y} = \bm{X} \beta + \bm{Z} u + \varepsilon,
\end{equation}
where $\bm{X}\in \mathbb{R}^{N\times p}$, $\bm{Z} \in \mathbb{R}^{N\times I}$, $\bm{Y}, \varepsilon \in \mathbb{R}^{N}$, the fixed effects $\beta \in \mathbb{R}^{p}$, and the random effects $u \in \mathbb{R}^{I}$. Let $N = 500$. The noise term is drawn independently from standard normal distribution, $\varepsilon_i \sim \mathcal{N}(0, \sigma_\varepsilon^2)$, with $\sigma_\varepsilon = 1$. Each row of $\bm{X}$ is independent and identically distributed from $\mathcal{N}(0_p, \Sigma_p)$ with $\Sigma_p = \mathbf{I}_p$. The design corresponding to the random effect is a balanced one-way ANOVA, with each cluster containing $n_s = N/I$ observations. The random effect is generated from a normal distribution such that $u \sim \mathcal{N}(0_I, \tau^2 \mathbf{I})$. We define the signal-to-noise ratio of the fixed effects to be $\beta^T\beta/\sigma_\varepsilon^2$. We generate $\beta$ by drawing $p$ independent uniform random variables in $[-1, 1]$ and then rescale the variables based on the signal-to-noise ratio. The results summarized below are based on 500 trials. 

Table \ref{res:null} presents the proportion of trials to reject the null hypothesis at $\alpha = 0.05$ using both the permutation null distribution and the asymptotic null distribution. The permutation null distribution is obtained from implementing 2,000 permutations of the cluster labels directly. We observe that the type I error rates are close to the nominal level of $0.05$ across different settings. In some cases (e.g., $p=1000$), the test appears slightly conservative, but overall the type I error rate is reasonably well-controlled at the nominal level under both permutation and asymptotic null distributions. 

\begin{table}[!ht] 
\centering
\caption{Type I error based on 500 experiments with random errors generated from a normal distribution, $N = 500$, $\tau^2=0$. Nominal level is set to $0.05$. }
\label{res:null}
\begin{tabular}{|c|c|ccc|ccc|} 
\hline
&& \multicolumn{3}{c|}{Permutation} & \multicolumn{3}{c|}{Asymptotic} \\
& & $p = 100$ & $p =500$ & $p = 1000$ &  $p = 100$ & $p =500$ & $p = 1000$ \\
\hline
\multirow{3}{*}{SNR $=1$} & $I = 5$   & 0.055  & 0.052 & 0.038  & 0.052  & 0.048  & 0.036\\
& $I = 10$   & 0.042  &  0.054&  0.048 & 0.038 & 0.052 & 0.048   \\
& $I = 20$   &  0.054 & 0.048 & 0.042  & 0.052  & 0.046  & 0.038  \\
 \hline
\multirow{3}{*}{SNR $=4$} & $I = 5$    &  0.048 & 0.036  & 0.042 & 0.046  & 0.032  & 0.042 \\
& $I = 10$   & 0.054  & 0.040 & 0.044  & 0.050  & 0.040  & 0.042\\
& $I = 20$   & 0.060 & 0.036 & 0.040  & 0.052  & 0.036 & 0.042\\
 \hline
 \multirow{3}{*}{SNR $=8$} & $I = 5$   & 0.062  & 0.054  & 0.052  & 0.062 &   0.052 & 0.048 \\
& $I = 10$   & 0.058  & 0.046 & 0.046  & 0.054  & 0.046  & 0.044 \\
& $I = 20$   &  0.052 & 0.050 & 0.048   & 0.054 & 0.052  & 0.048 \\
 \hline
\end{tabular}
\end{table}

We also include simulation setting where the random error is generated from non-normal data. Specifically, we generate $\varepsilon \sim t_3$ and then scale it to have unit variance. Each row of $\bm{X}$ is independent and identically distributed from $t_3$ as well. We find that even under non-normality, the test remains well-calibrated under the null, and the empirical size is very close to the nominal level for finite sample sizes. 

\begin{table}[!ht] 
\centering
\caption{Type I error based on 500 experiments with random errors generated from a $t_3$ distribution, $N = 500$, $\tau^2=0$. Nominal level is set to $0.05$. }
\label{res:null}
\begin{tabular}{|c|c|ccc|ccc|} 
\hline
&& \multicolumn{3}{c|}{Permutation} & \multicolumn{3}{c|}{Asymptotic} \\
& & $p = 100$ & $p =500$ & $p = 1000$ &  $p = 100$ & $p =500$ & $p = 1000$ \\
\hline
\multirow{3}{*}{SNR $=1$} & $I = 5$   & 0.042  & 0.056 & 0.060  & 0.044 & 0.054 & 0.060 \\
& $I = 10$    &  0.042 &  0.038 & 0.048 & 0.040 & 0.032  & 0.050\\
& $I = 20$    & 0.052  & 0.060 &  0.058 & 0.056  & 0.060  & 0.054 \\
 \hline
\multirow{3}{*}{SNR $=4$} & $I = 5$  & 0.048 & 0.036  & 0.042 & 0.046  & 0.032  & 0.042 \\
& $I = 10$ & 0.054  & 0.040  & 0.044  & 0.050  & 0.040  & 0.042 \\
& $I = 20$ & 0.050  & 0.040 &  0.050 & 0.052  & 0.040  & 0.050\\
 \hline
 \multirow{3}{*}{SNR $=8$} & $I = 5$  & 0.054 & 0.058  & 0.056  & 0.052  & 0.054  & 0.052  \\
& $I = 10$  & 0.048  & 0.038  & 0.042  &  0.044 & 0.040  & 0.038 \\
& $I = 20$  & 0.054  & 0.030  & 0.048  &  0.050 & 0.030  & 0.046 \\
 \hline
\end{tabular}
\end{table}

We compare the type I error control of our method to that of  \cite{law_inference_2023}, whose test targets a setting most closely related to ours. They propose an $F$-statistic to test whether a collection of random effects is zero in a high-dimensional linear mixed model. Their $F$-statistic is defined as: 
$$F_{EW} = \frac{||P_{\bm{Z}}(\bm{Y} - \bm{X}\hat{\beta}_{EW})||^2_2/\text{rank}(P_{\bm{Z}})}{||P^{\perp}_{\bm{Z}}(\bm{Y} - \bm{X}\hat{\beta}_{EW})||_2^2/\text{rank}(P^{\perp}_{\bm{Z}})},$$
where $P_{\bm{Z}}$ represents the projection onto the column space of $\bm{Z}$ and $P^{\perp}_{\bm{Z}}$ denotes the projection onto the orthogonal complement of the column space of $\bm{Z}$. 

Their method assumes that $\beta$ is a weakly sparse vector and estimates the regression coefficients, $\hat{\beta}_{EW}$, of the fixed effects via an exponential weighting procedure. However, under the null hypothesis of no random effect, their test statistic has difficulty controlling the nominal type I error rate if $\beta$ is not sufficiently sparse, even when the dimension of $\beta$ is moderately low. We compare the performance of their test statistic, $F_{EW}$, to our proposed test statistic under the null hypothesis, setting $p = 50$ and $N=500$. To estimate $\hat{\beta}_{EW}$, we follow Algorithm 1 from \cite{law2021inference} and obtain the necessary tuning parameters using cross-validation. We generate $\beta$'s so that the signal-to-noise ratio is set to the specified level. Type I error rates are based on 500 experiments and presented in Table \ref{res:null_com}, where $F_{EW}$ denotes the results using the $F$-statistic in \cite{law_inference_2023}. Critical value thresholds are obtained using 2000 permutations for both methods. We observe that, in these scenarios, the $F$-statistic does not adequately control the Type I error, whereas our proposed test statistic demonstrates effective control. 

\begin{table}[!ht] \centering
\caption{Comparison of Type I error for $F$-statistic ($F_{EW}$) and our proposed method ($Z_I)$. Type I errors are calculated based on 500 experiments, $p = 50$, $N=500$, and $\tau^2 = 0$. Nominal level is set to $0.05$. }
\label{res:null_com}
\begin{tabular}{|c|ccc|ccc|} 
\hline
& \multicolumn{3}{c|}{$F_{EW}$} & \multicolumn{3}{c|}{$Z_I$}\\
\cline{2-7}
 & $I = 4$ & $I = 5$ & $I = 10$ & $I = 4$ & $I = 5$ & $I = 10$\\
\hline
 {SNR $= 5$} & 0.42& 0.36 & 0.32 & 0.052 & 0.045 & 0.052 \\
\hline
{SNR $=8$} &  0.39 &  0.40 & 0.35 & 0.036  & 0.050   & 0.052   \\
\hline
{SNR $=15$} &  0.41 & 0.42  & 0.34 & 0.046  & 0.048  & 0.042  \\
 \hline
\end{tabular}
\end{table}

\subsection{Power Simulations}

We evaluate the power of our method under various scenarios, including high-dimensional linear mixed models with different random effect configurations, as well as when the data generating mechanism is non-linear. In all settings, the similarity graph $G_{\X}$ is a 20-MST constructed using Euclidean distance. We define power to be the proportion of trials to reject the null hypothesis at significance level $0.05$. To speed up computations, we consider 100 trials for each scenario. Power estimates are based on $2000$ permutations. 

\begin{itemize} 
\item \textit{Scenario 1: High-dimensional linear mixed model} \\
We generate data from the high-dimensional linear mixed model detailed in (\ref{eq:lmm}).  The random effect is generated from a normal distribution such that $u \sim \mathcal{N}(0_I, \Psi_I)$ with $\Psi_I = \tau^2I_P$.  The $\beta$'s are chosen such that signal-to-noise ratio of the fixed effect  ($\beta^T\beta/\sigma^2_\varepsilon$) is set to be five with $\sigma_\varepsilon=1$. The estimated power is reported in Table \ref{tab:power1}. We check $p = 200, 500, 1000$, with $\tau^2$ values varying from 0.5 to 2. 
\item \textit{Scenario 2: Correlated random effects} \\
 We generate data from the high-dimensional linear mixed model detailed in (\ref{eq:lmm}) but the random effect is generated from a normal distribution such that $u \sim \mathcal{N}(0_I, \Psi_I)$ with $\Psi_I = \tau^2 $ when $i=j$ and $\rho^{|i-j|}$ when $i \neq j$.  We set the signal-to-noise-ratio of the fixed effects to be five and set $I=5$, $p =500$, and $N=500$.  The estimated power for varying values of $\rho$ and $\tau^2$ is reported in Table \ref{tab:power2}. 
 \item \textit{Scenario 3: Multiple random effects} \\
 We generate data from the high-dimensional linear mixed model (\ref{eq:lmm}), but allow for interactions between $q$ of the fixed effects and the random effect. The design matrix $\mathbf{Z} = \text{block-diag}(\mathbf{Z}_1, \dots, \mathbf{Z}_I)$ is an $N \times I(q+1)$ matrix, where each $Z_s, s= 1, \hdots, I$ contains a column of ones (for the random intercept) and $q$ columns corresponding to fixed effects with random slopes. The random effects for each cluster are generated such that $u_s \sim \mathcal{N}(0_{q+1},\Psi_{q+1}),$ with $\Psi_{q+1} = \tau^2$ when $i=j$ and $\rho^{|i-j|}$ when $i \neq j$, for $s= 1, \hdots, I$. We set the signal-to-noise-ratio of the fixed effects to be five and fix $I=5$, $p =500$, and $N=500$. The estimated power for varying values of $q$ and $\tau^2$ is reported in Table \ref{tab:power3}. 
 \item \textit{Scenario 4: Non-linear mixed models} \\
We consider two non-linear mixed models. In both settings, the random effects design is a balanced one-way ANOVA design with either $I=5$ or $I=10$. The sample size is $N$ =  $500$.  We check the estimated power for $p = 200, 500, 1000$, with $\tau^2$ values varying from 1 to 2.5.  The first model is a non-linear transformation of the fixed effects only: 

\begin{equation} \label{eq:nonlin1}
    \bm{Y} = g(\bm{X} \beta) + \bm{Z} u + \varepsilon,
\end{equation} 
where $g(x_{ij}) = \log(|x_{ij}|)$. We generate $\beta$'s so that the signal-to-noise ratio is four. The results are presented in Table \ref{tab:power4a}. 

 The second model is a non-linear transformation of both fixed effects and random effects:
 \begin{equation} \label{eq:nonlin2}
    \bm{Y} = g(\bm{X} \beta + \bm{Z} u ) + \varepsilon,
\end{equation} 
where $g(x_{ij},u_i) =\frac{1+u_i}{1+\exp(-(x_{ij}\beta+u_i))}$. We generate $\beta$'s so that the signal-to-noise ratio is four. The results are presented in Table \ref{tab:power4b}. 
\end{itemize} 
 
Overall, in Tables \ref{tab:power1} through \ref{tab:power4b},  we observe that the number of rejections increases as $\tau^2$ increases. As the dimension $p$ increases, the power remains stable. The power also remains consistent for $I=5$ versus $I=10$. When the random effects are correlated, clusters will exhibit both within-cluster and inter-cluster correlation. The rationale of the testing procedure holds, but since observations from different clusters may still be close in $\Y$ space, this can dilute the within-cluster signal.  This is observed in Table \ref{tab:power2}: our testing procedure maintains reasonable power, as long as the correlation is not too strong. However, as the correlation increases, for the same $\tau^2$, we see that the power declines compared to the setting where the random effects are independent. Table \ref{tab:power3} demonstrates that our test can retain reasonable power even when each cluster has multiple correlated random effects, which are introduced to capture heterogeneity in the effects of several covariates. As the number of random slopes increases, the power declines for a fixed level of $\tau^2$. Notably, our test retains reasonable power even in the presence of non-linear relationships between the response, fixed effects, and random effects (Tables \ref{tab:power4a} and \ref{tab:power4b}).

While our method does not explicitly impose restrictions on the design matrix $\Z$, its rationale for detecting within-cluster similarity in $\Y$ implicitly relies on the presence of shared random intercepts to distinguish between the null and alternative hypotheses. For instance, if the random effects are present only as random slopes and not intercepts, the test tends to have low power. This is partially evident in Scenario 3, where power declines as additional random slopes are introduced ($q$ increases). This reduction in power can be attributed to the diminishing contrasts in intercepts, which becomes less detectable as the variation due to random slopes increases. 
 
\begin{table}[H] \centering
\caption{Reported power for varying values of $\tau^2$. Power is estimated as the proportion of trials (out of 100) that rejected the null hypothesis at significance level $0.05$. Data is generated under Scenario 1 for different values of $p$ and $I$ with $N=500$.}
\label{tab:power1}
\begin{tabular}{|c|c|ccccc|} 
\hline
& &  $\tau^2 = 0$ &  $\tau^2 = 0.5$ & $\tau^2 = 1$ & $ \tau^2= 1.5$ & $\tau^2=2$ \\
\hline
\multirow{3}{*}{$I=5$} 
& {$p = 200$}   & 0.05 & 0.58 & 0.79 & 0.84 & 0.94 \\
& {$p = 500$}  & 0.05 & 0.52 & 0.80 & 0.88 & 0.91  \\
& {$p = 1000$}  & 0.03 & 0.63 & 0.72 & 0.83 & 0.90  \\
 \hline
 \multirow{3}{*}{$I=10$} 
& {$p = 200$} & 0.06  & 0.52 & 0.78 & 0.86 & 0.94 \\
& {$p = 500$} &0.04 & 0.51 & 0.79 & 0.83 & 0.92  \\
& {$p = 1000$} & 0.04 & 0.45 & 0.76 & 0.81 & 0.94  \\
 \hline
\end{tabular}
\end{table}

\begin{table}[H] \centering
\caption{Reported power for varying values of $\tau^2$. Power is estimated as the proportion of trials (out of 100) that rejected the null hypothesis at significance level $0.05$.  Data is generated under Scenario 2 for different values of $\rho$, with $I=5$, $p=500$, and $N=500$.}
\label{tab:power2}
\begin{tabular}{|c|cccc|} 
\hline
&  $\tau^2 = 1$ & $\tau^2 = 1.5$ & $ \tau^2= 2$ & $\tau^2=2.5$ \\
\hline
{$\rho = 0.2$}  & 0.50 & 0.83 & 0.77 & 0.82 \\
{$\rho = 0.5$}  &  0.45 & 0.66 & 0.79 & 0.82  \\
{$\rho = 0.7$}  & 0.38 & 0.54 & 0.67 & 0.77 \\
 \hline
\end{tabular}
\end{table}

\begin{table}[H] \centering
\caption{Reported power for varying values of $\tau^2$. Power is estimated as the proportion of trials (out of 100) that rejected the null hypothesis at significance level $0.05$.  Data is generated under Scenario 3 for different values of $q$, with $I=5$, $p = 500$, and $N=500$.}
\label{tab:power3}
\begin{tabular}{|c|cccc|} 
\hline
&  $\tau^2 = 1$ & $\tau^2 = 1.5$ & $ \tau^2= 2$ & $\tau^2=2.5$ \\
\hline
{$q = 5$}  & 0.83 & 0.93 & 0.93 & 0.94 \\
{$q = 10$}  & 0.78 & 0.84 & 0.84 & 0.88   \\
{$q = 20$}  & 0.71 & 0.76 & 0.74 & 0.80  \\
 \hline
\end{tabular}
\end{table}

\begin{table}[H] \centering
\caption{Reported power for varying values of $\tau^2$. Power is estimated as the proportion of trials (out of 100) that rejected the null hypothesis at significance level $0.05$. Data is generated according to (\ref{eq:nonlin1}) under Scenario 4  for different values of $p$ with $N=500$. }
\label{tab:power4a}
\begin{tabular}{|c|c|cccc|c|} 
\hline
& &  $\tau^2 = 0.5$ & $\tau^2 = 1$ & $ \tau^2= 1.5$ & $\tau^2=2$ \\
\hline
\multirow{3}{*}{$I=5$} 
& {$p = 200$}  & 0.71 & 0.91 & 0.93 & 0.98 \\
& {$p = 500$}  & 0.82 & 0.87 & 0.96 & 0.99  \\
& {$p = 1000$}  &  0.85 & 0.92 & 0.95 & 0.97\\
 \hline
 \multirow{3}{*}{$I=10$} 
& {$p = 200$}  & 0.79 & 0.95 & 0.97 & 1.00 \\
& {$p = 500$}  &  0.70 & 0.93 & 0.98 & 1.00\\
& {$p = 1000$}  & 0.80 & 0.91 & 0.97 & 0.99\\
 \hline
 \end{tabular}
 \end{table}

 \begin{table}[H] \centering
\caption{Reported power for varying values of $\tau^2$. Power is estimated as the proportion of trials (out of 100) that rejected the null hypothesis at significance level $0.05$. Data is generated according to (\ref{eq:nonlin2}) under Scenario 4  for different values of $p$ with $N=500$.  }
\label{tab:power4b}
\begin{tabular}{|c|c|cccc|c|} 
\hline
& &  $\tau^2 = 1$ & $\tau^2 = 1.5$ & $ \tau^2= 2$ & $\tau^2=2.5$ \\
\hline
\multirow{3}{*}{$I=5$} 
& {$p = 200$}  & 0.75 & 0.82 & 0.78 & 0.87 \\
& {$p = 500$}  & 0.67 & 0.80 & 0.89 & 0.87 \\
& {$p = 1000$}  & 0.79 & 0.78 & 0.84 & 0.88 \\
 \hline
  \multirow{3}{*}{$I=10$} 
& {$p = 200$}  &  0.72 & 0.80 & 0.88 & 0.97\\
& {$p = 500$}  & 0.75 & 0.84 & 0.88 & 0.92 \\
& {$p = 1000$}  & 0.79 & 0.84 & 0.87 & 0.92\\
 \hline
 \end{tabular}
 \end{table}
    
\section{Data Application} \label{sec:app}

We consider a sorghum data application to evaluate the performance of our proposed test. Sorghum is the fifth most produced cereal and plays a significant role in sustainable and economically viable biofuel production \citep{chai2024root}. We analyze data from a study investigating how metabolite composition across three different sorghum genotypes influences biomass under varying water level conditions. The study design is as follows: sorgum were grown on a field located in Scottsbluff, Nebraska. The field was divided into six blocks. Each block contains two treatment (drought versus well-watered) and three sorghum genotypes. Root and leaf metabolites samples were obtained from two plants of the same genotype, under both treatments, and from different blocks in the field, for a total of $p =341$ metabolite covariates. Root and leaf sampling require excavation and ultimately destruction of the plant. There are 68 samples without missing values. Biomass data was log-transformed and metabolite data were scaled and also log-transformed. A 30-MST constructed from Euclidean distance was used as the similarity graph. We treat plant biomass as the response $\Y$,  the metabolite data and indicators for treatment, tissue sample, and genotype as fixed effects, 
and block as a random effect. We test whether the block effect has non-zero variance, which can inform decisions in the field about whether to include blocks in future experiments. The permutation p-value for testing the significance of block as a random effect is $0.0330$.   As a sanity check, we plot a heatmap of the covariate distances between the observations by tissue type and treatment. Visually, there appears to be a strong blocking effect; for example, in the well-water and drought leaf heatmaps, it appears that blocks 3 and 5 tend to be more similar, compared to observations outside the blocks. 

\begin{figure}[H]
\centering
\includegraphics[width=1\textwidth]{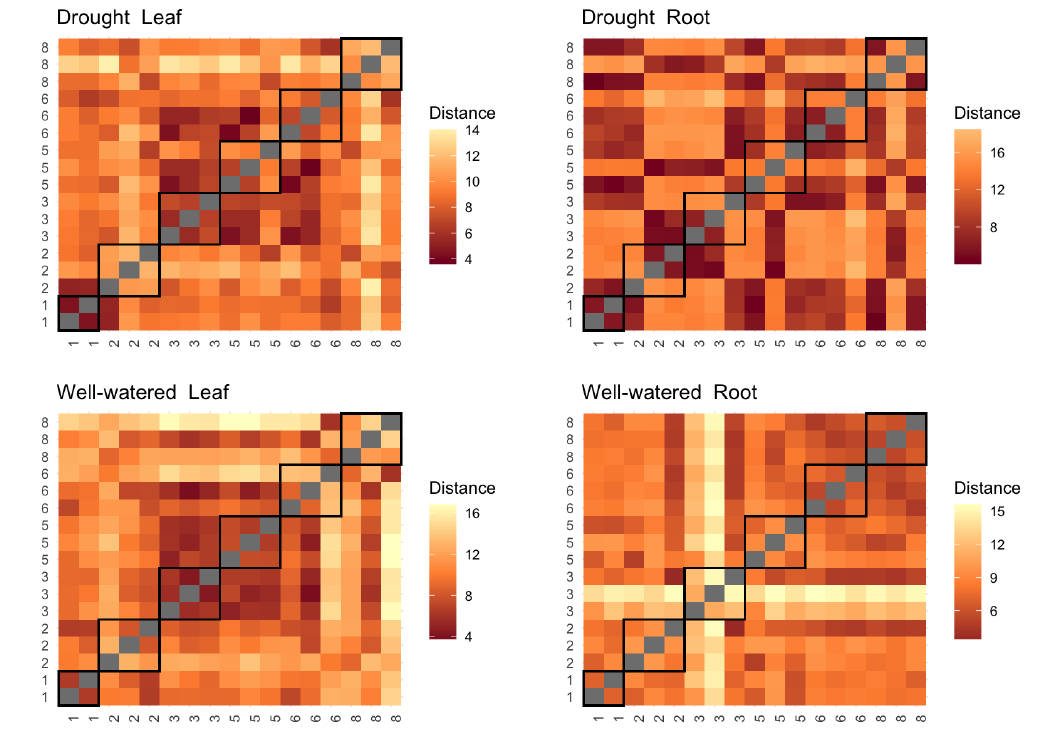}
\caption{Heatmap of distances between observations, subsetted by tissue type (leaf or root) and treatment type (drought or well-watered). Each black square represents the pairwise distances within the same block. The X and Y axis labels indicate the block to which the observations belongs to.}
\label{fig:Rep_8-1-2017_X}
\end{figure}


We also test for both block and genotype as random effects. The resulting $p$-value is less than $0.001$, indicating strong evidence that there are meaningful correlation structures in both blocks and genotypes. However, since only three genotypes are included in the study, it may be reasonable, depending on the research question objective, to include only block as a random effect in subsequent analyses.  

\section{Conclusion} \label{sec:conclusion}
We propose a new graph-theoretic test that utilizes both similarity information and rank-transformed data to assess whether incorporating group-level correlation meaningfully improves the prediction association of $\X$ on $\Y$. Our proposed test is model-free and can be utilized in a range of settings, including when the fixed effects are high-dimensional and/or the data-generating process is non-linear.  The test demonstrates reasonable power across a wide range of scenarios, including various configurations for random effects, and is well-calibrated under the null hypothesis of no random effect. The construction of the test statistic rests on the rationale that, if clustering exists, observations within-cluster would tend to be closer in both the $\Y$ and $\X$ space. The similarity graph, $G_{\X}$, captures proximity in $\X$ space, while the normalized ranks reflect relative position in $\Y$ space. An important factor influencing the test's performance is the density of the similarity graph. A denser graph (i.e., a larger $k$ in $k$-MST or $k$-NN) can incorporate additional similarity information into the test statistic, which can improve power. However, if $k$ is too large, the inclusion of  irrelevant similarity information can dilute the signal and ultimately reduce power. Thus, the choice of $k$ is application- and data- dependent, and selecting an optimal value remains an open problem. 

\bibliographystyle{apalike}
\bibliography{randomEffects}

\end{document}